\documentclass[12pt,preprintnumbers,superscriptaddress,nofootinbib]{revtex4}
\usepackage{multirow}
\usepackage{amssymb}
\usepackage{amsmath, graphicx}
\usepackage{dcolumn}
\usepackage{bm}
\usepackage{enumerate}
\usepackage{slashed}
\usepackage{epstopdf}
\usepackage[unicode]{hyperref}
\usepackage{csquotes}
\usepackage[usenames]{color}

\newcommand{\be}{\begin{equation}}
\newcommand{\ee}{\end{equation}}
\newcommand{\bea}{\begin{eqnarray}}
\newcommand{\eea}{\end{eqnarray}}
\newcommand{\ben}{\begin{enumerate}}
\newcommand{\een}{\end{enumerate}}
\newcommand{\bde}{\begin{widetext}}
\newcommand{\ede}{\end{widetext}}

\newcommand{\bc}{\begin{center}}
\newcommand{\ec}{\end{center}}

\usepackage{braket}

\begin{document}

\title{\boldmath Shaft inflation in Randall-Sundrum model}

\author{Ngo Phuc Duc Loc}
\email{locngo148@gmail.com}

\affiliation{Department of Physics and Astronomy, University of New Mexico, Albuquerque, NM 87131, USA}

\begin{abstract}
Shaft inflation is a model in which the inflaton potential approaches a plateau far from the origin, while it resembles chaotic inflation near the origin. Meanwhile, the Randall-Sundrum type II model (RSII) is an interesting extra-dimensional model to study cosmological phenomenology. In this paper, we study shaft inflation in the RSII model. We find that the predictions are in excellent agreement with observation. The fundamental five-dimensional Planck scale is found to be $M_5\simeq 10^{16}$ GeV, which is consistent with the lower bound $M_5\gtrsim 10^{9}$ GeV obtained from experimental Newtonian gravitational bound. This is an important result that can be used to explore further the implications of extra dimension in other contexts.
\end{abstract}
\maketitle

\tableofcontents

\section{Introduction}
Inflation is the leading paradigm to address the fine-tuned problem of the initial conditions of the Universe \cite{guth}, though the nature of the inflaton field responsible for driving inflation remains unknown. The latest results from Planck, WMAP, and BICEP/Keck observations suggest that there is not significant primordial tensor perturbations \cite{bicep}, which implies that concave-like inflaton potentials are more favorable.

From another aspect, the hierarchy problem, which is the vast discrepancy between the electroweak scale and the Planck scale \footnote{In popular terms, gravity is mysteriously much weaker than any other forces.}, motivated some possible solutions by modifying the structure of spacetime itself. Specifically, there are some models that introduce one or more extra spatial dimension(s) to explain the hierarchy problem; see Table \ref{models} for a summary \footnote{While extra spatial dimensional models are popular, it is also interesting to note that there exists an extra temporal dimensional model called Two-Time physics proposed by Itzhak Bars \cite{bars1,bars2}. In this model, there are 4 flat, non-compact spatial dimensions and 2 temporal dimensions. Interested readers can confer Ref. \cite{phong} for an inflationary scenario in this model.} . The basic idea is that gravity is much weaker than other forces because there are extra dimensions to leak gravity out. In other words, all the fields except gravity are confined on an effective $3+1$ dimensional spacetime called the brane, while in fact there exists a higher dimensional spacetime called the bulk. These proposals seem to be too speculative to be true, but in fact extra spatial dimensions naturally arise in a quantum gravity theory known as string theory \cite{joe}. A familiar argument for why these extra dimensions have not yet popped out in experiments is that they are tiny compact dimensions (such as RSI and ADD models), which means that we need a very high-energy scale and a very delicate instrument to detect, for example, a quick energy non-conservation process. An even more ambitious idea is that the extra dimensions are not small at all (such as RSII and DGP models), but still they do not cause any significant effects that can be easily detectable in a lab; for example, the RSII model predicts a modification to the Newton's gravitational law at small distance but there is a parameter space of the model that is still allowed within the experimental bound \cite{EPJC}. In this paper, we shall work on the Randall-Sundrum type II model (RSII) \cite{lisa2} as it is more interesting to study cosmological phenomenology in this model. Recently, it has even been shown that there can exist humanly traversable wormholes in this model \cite{maldacena}.

\begin{table}[h!]
	\caption{Comparisons between different models of extra spatial dimension(s). Note that the properties listed here are intrinsic features of the extra dimension(s) that were built into the model.}
\centering
\begin{tabular}{|c|c|c|c|c|c|} 
\hline
\multirow{2}{*}{Models} & \multirow{2}{*}{Number of extra dimensions} & \multicolumn{4}{|c|}{Properties of extra spatial dimensions} \\
\cline{3-6}
& & Compact & Non-compact & Flat & Curved \\
\hline
RSI\cite{lisa1}& 1 & \centering \checkmark & & & \checkmark 
\\
\hline
RSII\cite{lisa2}& 1 & & \checkmark & &\checkmark 
\\
\hline
ADD\cite{add1,add2,add3}& $\geq 2$ & \checkmark & & \checkmark & 
\\
\hline
DGP\cite{dgp} & 1 &  & \checkmark & \checkmark & 
\\
\hline
\end{tabular}
\label{models}
\end{table}

In this paper, we study the \textit{shaft inflation} model first introduced by Dimopoulos in Refs. \cite{kostas,kostas2} but now in the context of RSII scenario. This paper is organized as follows. In Sec. \ref{4D} we review the standard shaft inflation model in 4D spacetime. In Sec. \ref{RS2} we study shaft inflation in RSII model. We compare the results of two models with each other and with observations in Sec. \ref{comparisons}. Conclusions are in Sec. \ref{conclu}. Natural units in which $\hbar=c=k_B=1$ are used throughout this paper.

\section{Shaft inflation in 4D spacetime}\label{4D}
\begin{figure}[h!]
\centering
\includegraphics[scale=1.2]{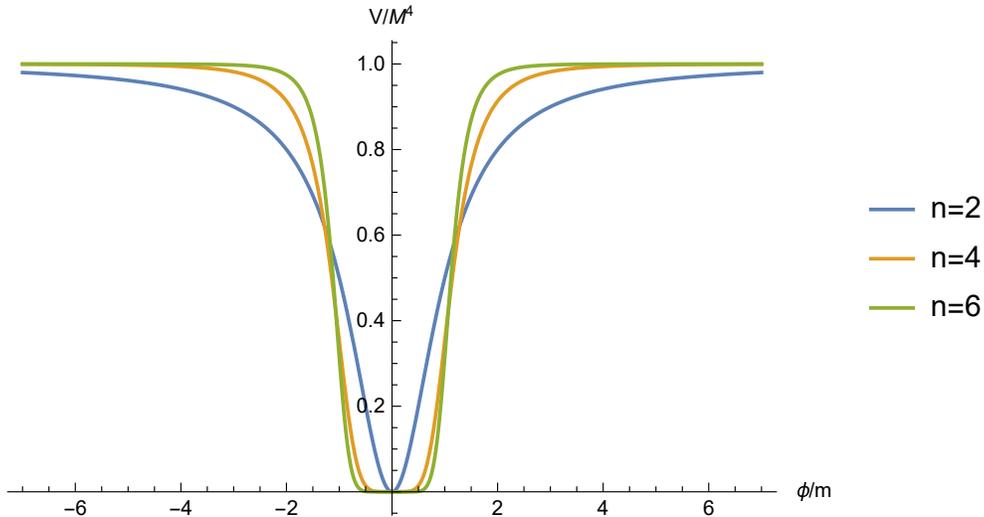}
\caption{The shaft inflaton potential with $n=2,4,6$.}
\label{potential_figure}
\end{figure}

For future comparisons, here we review the standard shaft inflation model in the usual 3+1 dimensional spacetime \cite{kostas}. The inflaton potential  is
\begin{equation}\label{potential}
V(\phi)=M^4\phi^{2(n-1)}(\phi^n+m^n)^{2(1-n)/n},
\end{equation}
where $M$ is the energy scale of inflation, $m$ is roughly the threshold energy scale for the inflaton field to stop slow-roll, and $n\geq 2$ is an integer. The main feature is that when $\phi\gg m$ the potential approaches a constant value (plateau), while when $\phi\ll m$ the potential looks similar to a monomial chaotic inflation. The potential is plotted in Figure \ref{potential_figure}. For $n=2$, the potential is realized in S-dual superstring inflation with $m\simeq m_P/4$ \cite{superstring} and in radion assisted gauge inflation with $m\simeq 10^{-1.5}m_P$ \cite{radion}, where 
$m_P\simeq 2.435\times 10^{18}$ GeV is the reduced Planck mass. We will choose $m\simeq 10^{18}$ GeV in our calculations as inflation generally happens when the inflaton field is above this energy scale in large-field inflation models. (See also \cite{gas} for a study of shaft inflaton potential in a warm inflation scenario.)

\subsection{Dynamics}
\begin{figure}[h!]
\centering
\includegraphics[scale=1.2]{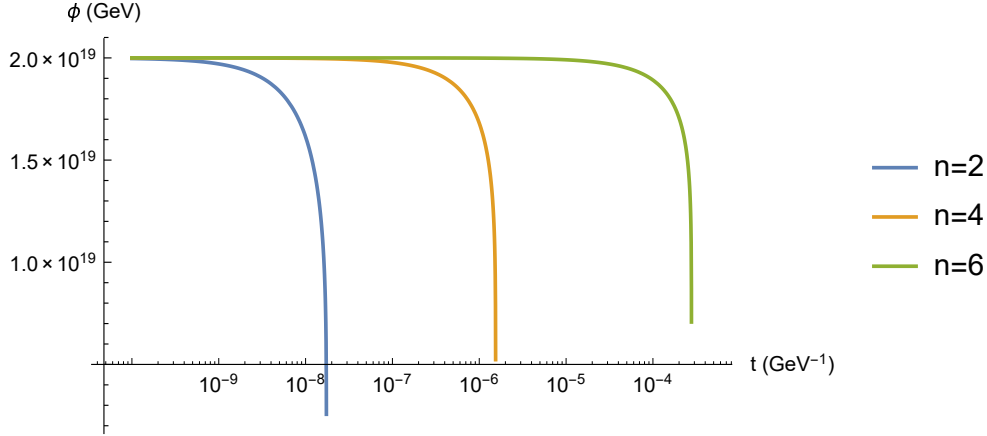}
\caption{Evolution of the inflaton field in the standard 4D spacetime model with $n=2,4,6$. The parameters are $M\simeq 10^{15}$ GeV, $M_4\simeq 10^{19}$ GeV, $m\simeq 10^{18}$ GeV, and $\phi_i\equiv 2\times 10^{19}$ GeV. }
\label{field}
\end{figure}

The Klein-Gordon equation in the slow-roll approximation is
\begin{equation}
\dot{\phi}\simeq-\frac{V'}{3H}=-\frac{M_4}{\sqrt{24\pi}}\frac{V'}{\sqrt{V}}\simeq-\frac{M_4M^2m^n(n-1)}{\sqrt{6\pi}}\frac{1}{\phi^{n+1}},
\end{equation}
where $M_4$ is the familiar 4D Planck scale. Note that we said earlier that inflation happens when $\phi>m$, so we will use this approximation frequently in our calculations. The solution of this equation is
\begin{equation}\label{phi(t)4D}
\phi(t)=\left[(2+n)\left(\frac{m^nM^2M_4(1-n)}{\sqrt{6\pi}}t+\frac{\phi_i^{2+n}}{2+n}\right)\right]^{1/(2+n)},
\end{equation}
where $\phi_i$ is the initial value of the inflaton field. The evolution of the inflaton field is sketched in Figure \ref{field}. Compare Fig. \ref{field} and Fig. \ref{potential_figure}, we see an expected pattern: as $n$ increases, the inflaton field rolls more slowly because the top of the potential becomes flatter.

Meanwhile, the Friedman equation in the slow-roll approximation is
\begin{equation}
H^2\equiv\left(\frac{\dot{a}(t)}{a(t)}\right)^2\simeq\frac{8\pi}{3M_4^2}V\simeq \frac{8\pi M^4}{3M_4^2},
\end{equation}
whose solution is
\begin{equation}\label{scale_factor_4D}
a(t)=a_i\exp\left(\sqrt{\frac{8\pi}{3}}\frac{M^2}{M_4}t\right),
\end{equation}
where $a_i$ is the initial scale factor.

\subsection{Observables}

The slow-roll parameters are
\begin{equation}
\epsilon=\frac{M_4^2}{16\pi}\left(\frac{V'}{V}\right)^2=\frac{M_4^2m^{2n}(n-1)^2}{4\pi}\frac{1}{\phi^2(\phi^n+m^n)^2},
\end{equation}
\begin{equation}
\eta=\frac{M_4^2}{8\pi}\left(\frac{V''}{V}\right)=\frac{M_4^2m^n(n-1)}{4\pi}\frac{[m^n(2n-3)-(1+n)\phi^n]}{\phi^2(\phi^n+m^n)^2}.
\end{equation}
The condition to end inflation is $|\eta|\simeq 1$, which implies
\begin{equation}\label{phi_f}
\phi_f=\left(\frac{M_4}{\sqrt{8\pi}}\right)^{2/(n+2)}\left[2(n^2-1)m^n\right]^{1/(n+2)}.
\end{equation}
The number of e-folds is
\begin{equation}
N=\frac{8\pi}{M_4^2}\int_{\phi_f}^\phi\frac{V}{V'}d\phi\simeq\frac{4\pi}{M_4^2m^n(n-1)(n+2)}(\phi^{n+2}-\phi_f^{n+2}).
\end{equation}
From this and Eq. \ref{phi_f}, we get
\begin{equation}\label{phi_i}
\phi(N)=\left[\frac{M_4^2m^n(n-1)[(n+2)N+n+1]}{4\pi}\right]^{1/(n+2)}.
\end{equation}
The scalar spectral index is
\begin{align}
n_s&=1-6\epsilon+2\eta\\
&=1-\frac{M_4^2m^n(n-1)}{2\pi}\left[\frac{nm^n+(1+n)\phi^n}{\phi^2(\phi^n+m^n)^2}\right]\\
&\simeq 1-\frac{2(n+1)}{(n+2)N+n+1},\label{ns}
\end{align}
where we used Eq. \ref{phi_i}. 
The tensor-to-scalar ratio is
\begin{align}
r&=16\epsilon\\
&\simeq \frac{4M_4^2m^{2n}(n-1)^2}{\pi}\frac{1}{\phi^{2(n+1)}}\\
&=4\left(\frac{\sqrt{\pi}m}{M_4}\right)^{2n/(n+2)}(n-1)^2\left[\frac{4}{(n-1)[(n+2)N+n+1]}\right]^{2(n+1)/(n+2)},\label{r}
\end{align}
where we used Eq. \ref{phi_i}.

To calculate the running spectral index, we need another slow-roll parameter
\begin{equation}
\xi=\frac{M_4^4}{64\pi^2}\frac{V'V'''}{V^2}.
\end{equation}
The running spectral index is then
\begin{align}
\frac{dn_s}{d\ln k}&=16\epsilon\eta-24\epsilon^2-2\xi\\
&=-\frac{m^{2n}M_4^4(n-1)^2}{8\pi^2}\frac{\left[2nm^{2n}+m^n(2+3n+n^2)\phi^n+(2+3n+n^2)\phi^{2n}\right]}{\phi^4(m^n+\phi^n)^4}\\
&\simeq -\frac{m^{2n}M_4^4(n-1)^2(2+3n+n^2)}{8\pi^2}\frac{1}{\phi^{2(n+2)}}\\
&=-\frac{2(n^2+3n+2)}{\left[(n+2)N+n+1\right]^2}\sim-\frac{2}{N^2}.
\end{align}
With $N\sim 50-60$ we get $dn_s/d\ln k\sim 10^{-4}$, so the running spectral index is very small which is compatible with observations in Ref.  \cite{planck}.

We will discuss the predictions of shaft inflation in 4D spacetime in Sec. \ref{comparisons}.

\section{Shaft inflation in RSII model}\label{RS2}
According to the RSII model, the inflaton field is confined on the brane and only the (effective) gravitational background is modified. This reflects the fact that gravity can work in 4+1 dimensional spacetime (in the bulk) but matter fields only exist in the usual 3+1 dimensional spacetime (on the brane). As a consequence, the Friedman equation on the brane is modified as $H^2\propto \rho+\rho^2$. In the high-energy limit when inflation happens, we have $\rho\simeq V(\phi)$ and the quadratic term is dominant, so that $H^2\propto \rho^2$ during inflation. At late times such as the radiation or matter dominated eras, the quadratic term decays much faster than the linear term, so the usual Friedmann equation $H^2\propto\rho$ is recovered after inflation ends and sets the stage for the subsequent usual Big Bang cosmology. The modification of the Hubble rate at early times in turn affects the predictions of inflationary models. We follow a common assumption that the extra dimension does not expand nor contract; it only serves the role of modifying the effective gravitational background on the brane. Here, we study the predictions of the shaft inflaton potential (Eq. \ref{potential}) in RSII model. The general framework of inflation in RSII model is discussed in \cite{roy} and references therein.

\subsection{Dynamics}
The Klein-Gordon equation in the slow-roll approximation is
\begin{equation}
\dot{\phi}\simeq-\frac{V'}{3H}=-\frac{M_5^3}{4\pi}\frac{V'}{V}\simeq-\frac{M_5^3m^n(n-1)}{2\pi}\frac{1}{\phi^{n+1}},
\end{equation}
where $M_5$ is the five-dimensional Planck scale. The parameter $M_5$ can be thought of as the fundamental energy scale of gravity in the bulk, whereas the usual 4D Planck scale $M_4$ is just an effective energy scale of gravity on the brane. The solution of this equation is
\begin{equation}\label{phi(t)RS2}
\phi(t)=\left[(2+n)\left(\frac{m^nM_5^3(1-n)}{2\pi}t+\frac{\phi_i^{2+n}}{2+n}\right)\right]^{1/(2+n)},
\end{equation}
where $\phi_i$ is the initial value of the inflaton field.

Meanwhile, the Friedman equation in the slow-roll approximation is
\begin{equation}
H^2\equiv\left(\frac{\dot{a}(t)}{a(t)}\right)^2\simeq \frac{16\pi^2}{9M_5^6}V^2\simeq\frac{16\pi^2M^8}{9M_5^6},
\end{equation}
whose solution is 
\begin{equation}\label{scale_factor_RS2}
a(t)=a_i\exp\left(\frac{4\pi M^4}{3M_5^3}t\right),
\end{equation}
where $a_i$ is the initial scale factor.

\subsection{Observables}

Given our potential in Eq. \ref{potential}, the slow-roll parameters are
\begin{equation}
\epsilon=\frac{3M_5^6}{16\pi^2}\frac{V'^2}{V^3}=\frac{3m^{2n}M_5^6(n-1)^2}{4M^4\pi^2}\frac{1}{\phi^{2n}(\phi^n+m^n)^{2/n}},
\end{equation}
\begin{equation}
\eta=\frac{3M_5^6}{16\pi^2}\frac{V''}{V^2}=\frac{3m^nM_5^6(n-1)}{8M^4\pi^2}\frac{[m^n(2n-3)-(1+n)\phi^n]}{\phi^{2n}(\phi^n+m^n)^{2/n}}.
\end{equation}
From the condition for inflation to end $|\eta|\simeq 1$, we obtain the value of the inflaton field at the end of inflation
\begin{equation}\label{phi_f_RS2}
\phi_f=\left(\frac{3m^nM_5^6(n^2-1)}{8M^4\pi^2}\right)^{1/(n+2)}.
\end{equation}
The number of e-folds is
\begin{equation}
N=\frac{-16\pi^2}{3M_5^6}\int_\phi^{\phi_f}\frac{V^2}{V'}d\phi\simeq\frac{8\pi^2M^4}{3M_5^6m^n(n-1)(n+2)}(\phi^{n+2}-\phi_f^{n+2}).
\end{equation}
From this and Eq. \ref{phi_f_RS2} we get
\begin{equation}\label{phi_i_RS2}
\phi(N)=\left[\frac{3M_5^6m^n(n-1)[(n+2)N+(n+1)]}{8\pi^2M^4}\right]^{\frac{1}{n+2}}.
\end{equation}
The scalar spectral index is
\begin{align}
n_s&=1-6\epsilon+2\eta\\
&=1-\frac{3m^nM_5^6(n-1)}{4M^4\pi^2}\left[\frac{m^n(4n-3)+(1+n)\phi^n}{\phi^{2n}(\phi^n+m^n)^{2/n}}\right]\\
&\simeq 1-\frac{2(n+1)}{(n+2)N+n+1},\label{ns_RS2}
\end{align}
where we used Eq. \ref{phi_i_RS2}. The tensor-to-scalar ratio is
\begin{align}
r&=24\epsilon\\
&\simeq\frac{18m^{2n}M_5^6(n-1)^2}{M^4\pi^2}\frac{1}{\phi^{2(n+1)}}\\
&=18\left(\frac{\pi mM^2}{M_5^3}\right)^{2n/(n+2)}(n-1)^2\left[\frac{8}{3(n-1)[(n+2)N+n+1]}\right]^{2(n+1)/{(n+2)}},\label{r_RS2}
\end{align}
where we used Eq. \ref{phi_i_RS2}. Compare Eq. \ref{ns_RS2} with Eq. \ref{ns}, we see that the scalar spectral index in two models are the same. Compare Eq. \ref{r_RS2} with  Eq. \ref{r},  we see that the tensor-to-scalar ratio in RSII model now also depends on the energy scale of inflation $M$.

Naively one can expect that the running spectral index in RSII model would not be different from the standard 4D case as we saw that the scalar spectral index in two models are the same. But let's also check this more explicitly. To calculate the running spectral index, we need another slow-roll parameter
\begin{equation}
\xi=\left(\frac{3M_5^6}{16\pi^2}\right)^2\frac{V'V'''}{V^4}.
\end{equation}
The running spectral index is then (cf. Ref. \cite{lidsey} for the general formula of running spectral index in RSII model)
\begin{align}
\frac{dn_s}{d\ln k}&=16\epsilon\eta-18\epsilon^2-2\xi\\
&= -\frac{9m^{2n}M_5^{12}(n-1)^2}{32M^8\pi^4}\frac{\left[2nm^{2n}(4n-3)+3m^n(3n^2+n-2)\phi^n+(n^2+3n+2)\phi^{2n}\right]}{\phi^{4n}(m^n+\phi^n)^{4/n}}\\
&\simeq -\frac{9m^{2n}M_5^{12}(n-1)^2(n^2+3n+2)}{32M^8\pi^4}\frac{1}{\phi^{2(n+2)}}\\
&=-\frac{2(n^2+3n+2)}{\left[(n+2)N+n+1\right]^2}\sim-\frac{2}{N^2}.
\end{align}
So indeed the running spectral indices are the same in two models.

Another important observable is the amplitude of scalar perturbation:
\begin{align}
A_s&\simeq\frac{1024\pi^4}{81M_5^{18}}\frac{V^6}{V'^2}\\
&= \frac{256\pi^4M^{16}}{81M_5^{18}m^{2n}(n-1)^2}\frac{1}{\phi^{6-8n}(m^n+\phi^n)^{6-8/n}}\\
&\simeq \frac{256}{81}\left[\frac{\pi^2M^{4(n+3)}}{(n-1)m^nM_5^{3(n+4)}}\right]^{2/(n+2)}\left[\frac{3((n+2)N+n+1)}{8}\right]^{2(n+1)/(n+2)}.\label{A_s}
\end{align}

We will discuss the results of shaft inflation in RSII model in the next section.

\section{Comparisons}\label{comparisons}
\begin{figure}[h!]
\centering
\includegraphics[scale=1.2]{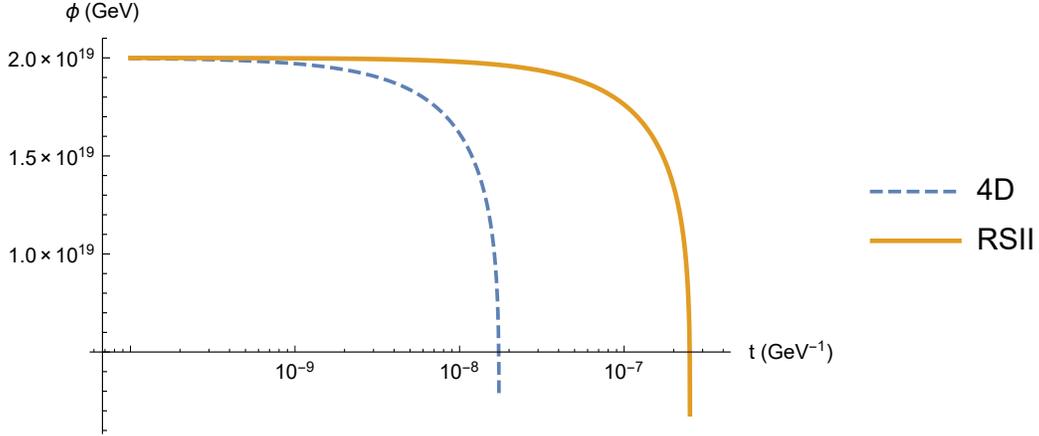}
\caption{Evolution of the inflaton field in 4D spacetime model versus RSII model with $n=2$. The parameters are $M\simeq 10^{15}$ GeV, $m\simeq 10^{18}$ GeV, $M_4\simeq 10^{19}$ GeV, $M_5\simeq 10^{16}$ GeV, and $\phi_i\equiv 2\times 10^{19}$ GeV.}
\label{field2}
\end{figure}

Let us first compare the dynamics of shaft inflation in two models. The evolution of the inflaton field in 4D spacetime (Eq. \ref{phi(t)4D}) and in RSII model (Eq. \ref{phi(t)RS2}) is plotted in Fig. \ref{field2} (for the case $n=2$). We see that the inflaton field rolls more slowly in RSII model. This feature is expected in RSII's cosmology as the Friedman equation is quadratic in energy density, which means that there is more friction for the evolution of the inflaton field according to the Klein-Gordon equation.

\begin{figure}[h!]
\centering
\includegraphics[scale=1.2]{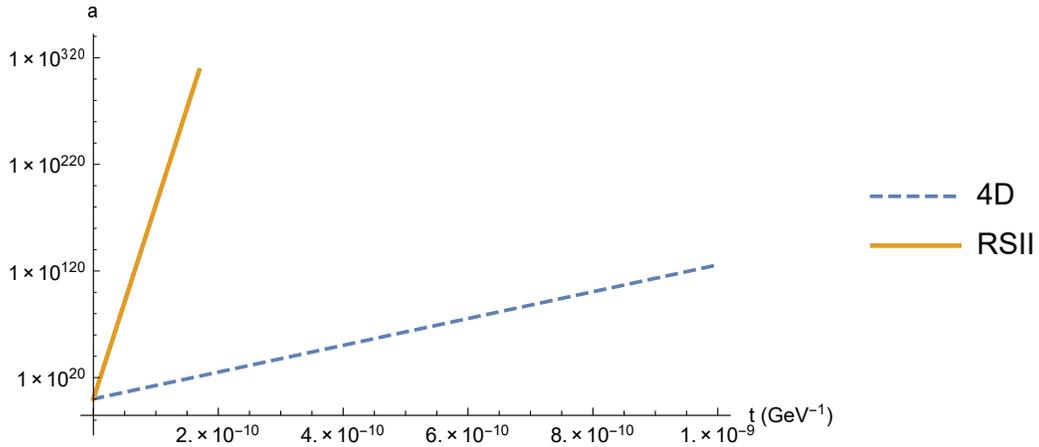}
\caption{Comparison of the scale factor between the 4D spacetime model and the RSII model. The parameters are $M\simeq 10^{15}$ GeV, $M_4\simeq 10^{19}$ GeV, $M_5\simeq 10^{16}$ GeV, and $a_i\equiv 1$.}
\label{scalefactor}
\end{figure}

Next, let's compare the evolution of the scale factor in two models. The scale factor in 4D spacetime (Eq. \ref{scale_factor_4D}) and in RSII model (Eq. \ref{scale_factor_RS2}) is plotted in Fig. \ref{scalefactor}. We see that the scale factor in RSII model grows more rapidly than the 4D case. This is again because of the fact that $H^2\sim\rho^2$ during inflation in RSII model \footnote{Note that the scale factors increase exponentially during inflation but we used the log scale in the plot for ease of comparison.}.

Now, we compare the observables of two models with each other and with observations. The typical energy scale of inflation is $M\simeq 10^{15}$ GeV and assuming that $m\simeq 10^{18}$ GeV, we can find the value of $M_5$ by demanding that the result in Eq. \ref{A_s} match the observed value $A_s^{Planck}\sim 2.0989\times 10^{-9}$ \cite{planck}. It is found that $M_5$ is typically of order $M_5\sim 10^{16}$ GeV (it depends only weakly on $n$ and $N$); this is consistent with a previously found lower bound of $M_5$ of order $10^9$ GeV \cite{EPJC} \footnote{There can also exist an upper bound if one wants to use the model to address other problems such as baryogenesis \cite{loc} but we will not consider it here.} . So, we find the results in Table \ref{4DvsRS2}. From Table \ref{4DvsRS2}, we see that  the RSII model predicts larger tensor-to-scalar ratio than the 4D case. This is more or less an expected feature of inflation in RSII model (for example, in Ref. \cite{EPJC} it was found that the tensor-to-scalar ratio in RSII model is larger than that in 4D spacetime for the concave inflaton potential of the form $V\propto\phi^n$ with $0<n<1$).

\begin{table}[h!]
\caption{Comparison of observables between the 4D spacetime model and the RSII model.}
\centering
\begin{tabular}{|c|c|>{\centering}p{1cm}|>{\centering}p{1.8cm}|c|c|c|} 
\hline
Models & Observables & N & $n=2$ & $n=4$ & $n=6$ & $n=8$\\
\hline
\multirow{4}{*}{4D} & \multirow{2}{*}{$n_s$} & 50 & 0.970 & 0.967 & 0.966 & 0.965\\
\cline{3-7}
& & 60 & 0.975 & 0.973 & 0.971 & 0.970\\
\cline{2-7}
& \multirow{2}{*}{$r$} & 50 & 0.002 & $4.2\times 10^{-4}$ & $1.4\times 10^{-4}$ & $6\times 10^{-5}$\\
\cline{3-7}
& & 60 & 0.001 & $3.1\times 10^{-4}$ & $10^{-4}$ & $4.4\times 10^{-5}$\\
\hline
\multirow{4}{*}{RSII} & \multirow{2}{*}{$n_s$} & 50 & 0.970 & 0.967 & 0.966 & 0.965\\
\cline{3-7}
& & 60 & 0.975 & 0.973 & 0.971 & 0.970\\
\cline{2-7}
& \multirow{2}{*}{$r$} & 50 & 0.058 & 0.018  & 0.005 & 0.002 \\
\cline{3-7}
& & 60 & 0.041 & 0.012 & 0.003 & 0.001\\
\hline
\end{tabular}
\label{4DvsRS2}
\end{table}

\begin{figure}[h!]
\centering
\includegraphics[scale=0.7]{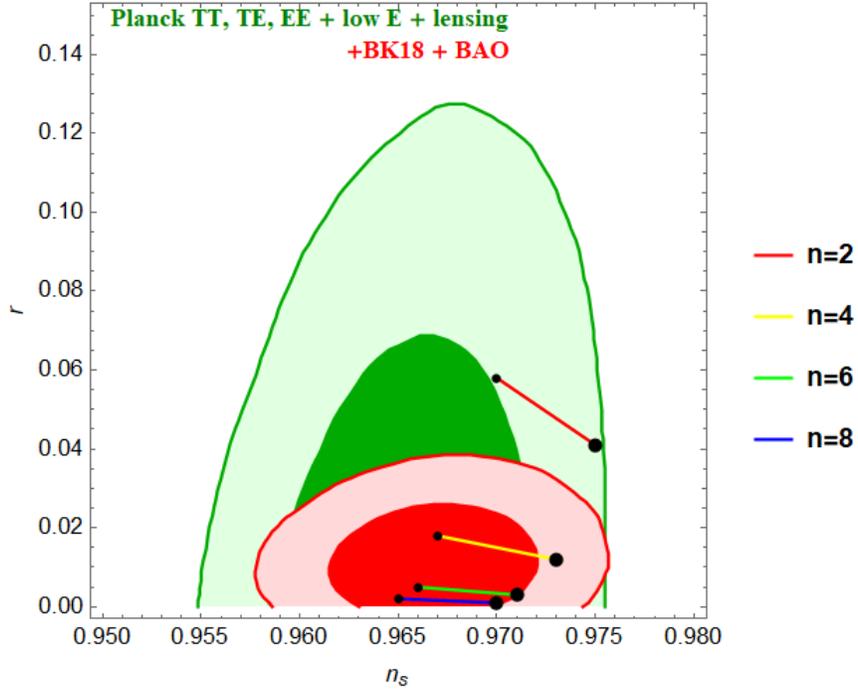}
\caption{Comparison between predictions of shaft inflation in RSII model and observation. Small (Large) black dots correspond to $N=50$ ($N=60$), respectively.  The parameters are $M\simeq 10^{15}$ GeV and $M_5\simeq 10^{16}$ GeV. }
\label{planck_figure}
\end{figure}

Comparison between predictions of RSII model and observations in Ref. \cite{bicep} is shown in Fig. \ref{planck_figure}. We see that the results are in excellent agreement with observation for various values of $n$. Unlike the 4D case, the RSII model predicts larger tensor-to-scalar ratios that can be detected in future precise observations.

\newpage

\section{Conclusions}\label{conclu}
In this paper, we studied shaft inflation in RSII model. We then compared the results with the standard 4D spacetime case and with observations. From the dynamical perspective, we obtained an intuitively expected feature that the inflaton field rolls more slowly in RSII model due to the modification of the Friedman equation during inflation. Specifically, the Hubble rate squared is proportional to the quadratic energy density and hence it generates more friction in the Klein-Gordon equation. The rapid growth of the scale factor in RSII model ensures that it can easily generate a sufficient number of e-folds to explain the horizon and flatness problems. From the more important observational perspective, the predictions of shaft inflation in RSII model are in excellent agreement with observation, though the case  $n=2$ is ruled out when taking into account the BICEP/Keck result. The fundamental five-dimensional Planck scale was found to be $M_5\simeq 10^{16}$ GeV, which is consistent with the previously found lower bound of this parameter $M_5\gtrsim 10^9$ GeV obtained from  Newtonian gravitational bound \cite{EPJC}. The value of $M_5$ that we found is close to the GUT scale and the energy scale of inflation which is of order $10^{15}$ GeV. It is an important prediction and can be used to explore further the effects of the extra dimension in other contexts.

\section*{Acknowledgment}
The author is grateful for financial support from the Department of Physics and Astronomy at UNM through the Origins of the Universe Award.

\end{document}